\documentclass{agujournal2019}
\usepackage{url} 
\usepackage{lineno}
\usepackage{graphicx}
\usepackage{enumerate}

\draftfalse

\journalname{JGR: Space Physics}

\begin{document}

\title{High-altitude polar NM with the new DAQ system as a tool to study details of the cosmic-ray induced nucleonic cascade }

%
%




\authors{Markus Simil\"a\affil{1,2}, Ilya Usoskin\affil{1,2}, Stepan Poluianov\affil{1,2}, Alexander Mishev\affil{1,2},
 Gennady A. Kovaltsov\affil{3}, Du Toit Strauss\affil{4}}

\affiliation{1}{Space Physics and Astronomy Research Unit, University of Oulu, Finland}
\affiliation{2}{Sodankyl\"a Geophysical Observatory, University of Oulu, Finland}
\affiliation{3}{Ioffe Physical-Technical Institute RAS, St. Petersburg, Russia}
\affiliation{4}{Center for Space Research, North-West University, Potchefstroom, South Africa}


\correspondingauthor{Ilya Usoskin}{ilya.usoskin@oulu.fi}



\begin{keypoints}
\item New data-acquisition (DAQ) systems, digitizing all pulses at a 2-MHz sampling rate, is installed and tested at DOMC/DOMB neutron monitors.
\item Several branches are identified in the pulse parameters, corresponding to different processes.
\item The new DAQ system allows studying inter-cascade, intra-cascade and instrumental response separately from the same dataset.
\end{keypoints}


\begin{abstract}
A neutron monitor (NM) is, since the 1950s, a standard ground-based detector whose count rate reflects cosmic-ray variability.
The worldwide network of NMs forms a rough spectrometer for cosmic rays.
Recently, a brand-new data acquisition (DAQ) system has been installed on the DOMC and DOMB NMs, located at the Concordia research
 station on the Central Antarctic plateau.
The new DAQ system digitizes, at a 2-MHz sampling rate, and records all individual pulses corresponding to secondary particles
 in the detector.
An analysis of the pulse characteristics (viz. shape, magnitude, duration, waiting time) has been performed, and several
 clearly distinguishable branches were identified:
 (A) corresponding to signal from individual secondary neutrons;
 (B) representing the detector's noise;
 (C) double pulses corresponding to the {shortly separated nucleons of the same} atmospheric cascades;
 (D) very-high multiple pulses which are likely caused by atmospheric muons;
 and (E) double pulses potentially caused by contamination of the neighbouring detector.
An analysis of the waiting time distributions has revealed two clearly distinguishable peaks: peak (I) at about 1 msec being
 related to the intra-cascade diffusion and thermalisation of secondary atmospheric neutrons; and peak (II) at 30\,--\,1000
 msec corresponding to individual atmospheric cascades.
This opens a new possibility to study spectra of cosmic-ray particles in a single location as well as details of the cosmic-ray induced
 atmospheric cascades, using the same dataset.
\end{abstract}


\section{Introduction}

A neutron monitor (NM) is a standard ground-based detector to monitor cosmic-ray variability in
 the near-Earth environment \cite{shea_SSR_00,vainio09,usoskin_gil_17}.
The design of the NM was developed in 1957 (called IGY --- International Geophysical Year) and
 improved in 1964 (called NM64), and since then it is used as a standard detector \cite{simpson58,simpson00,stoker09}.
NMs record primarily the secondary nucleonic component (mostly neutrons) of the cosmic-ray induced atmospheric cascade
 with a small fraction of counts caused by muons.
Its count rate is defined by the flux of primary (impinging on the top of the atmosphere) cosmic rays,
 as a combination of the cosmic-ray energy spectrum, detector's yield function and geomagnetic rigidity cutoff
 \cite{clem00,mishev20}.
The NM is an energy-integrating detector, with the effective energy ranging from about 12 GeV for polar NMs
 to 35 GeV for equatorial ones \cite{asvestari_JGR_17}.
The sensitivity of NMs to low-energy cosmic rays is highest in polar regions (low or no geomagnetic shielding)
 and high altitudes (lower atmospheric shielding) and decreases towards equatorial latitudes.
The worldwide network of NMs can act as a giant spectrometer able to roughly estimate the spectrum of
 both galactic cosmic rays \cite<e.g.,>[]{dorman04} and relativistic solar protons \cite<e.g.,>[]{duggal79,mishev14}.
Along with the count rate, the multiplicity of NM counts (the average number of pulses within a short time interval)
 is sometimes studied \cite{dorman04,balabin11} as a rough index of the spectral hardness of cosmic rays.
Sometimes NM are accompanied by separate muon detectors to measure high-energy cosmic rays.

This work is focused at two mini-NMs, DOMC and DOMB, located at the Concordia Antarctic Research station
 on top of Dome C, Central Antarctic plateau (75$^\circ$06'S, 123$^\circ$23'E, 3233 m above sea level) \cite{poluianov15}.
They are ones of the most sensitive NMs to lower energy cosmic rays (including solar energetic particles)
 thanks to the highly elevated polar location.
Each NM has one BF$_3$-filled detector surrounded by reflecting and moderating layers of polyethylene.
In addition, DOMC has a layer of lead serving as a neutron producer to increase the detector efficiency.
DOMB has no lead layer and therefore has lower efficiency than DOMC, but is more sensitive to low-energy
 secondary cosmic-ray particles \cite{vashenyuk07}.
Recently, those instruments got a major upgrade of the data-acquisition system (DAQ).
Traditionally, a standard NM records only the count rate of cosmic rays, while the new electronics of DOMC
 and DOMB digitizes individual detector pulses with a sub-microsecond precision \cite<2-MHz sampling rate, see>[]{strauss20}.
In this work, we study new opportunities provided by the DAQ upgrade, by using
 the statistic of recorded pulses and show that it allows one to study details of
 the cosmic-ray induced atmospheric cascade with instruments like DOMB and DOMC.

\begin{figure}[t]
\centering
\includegraphics[width=\columnwidth]{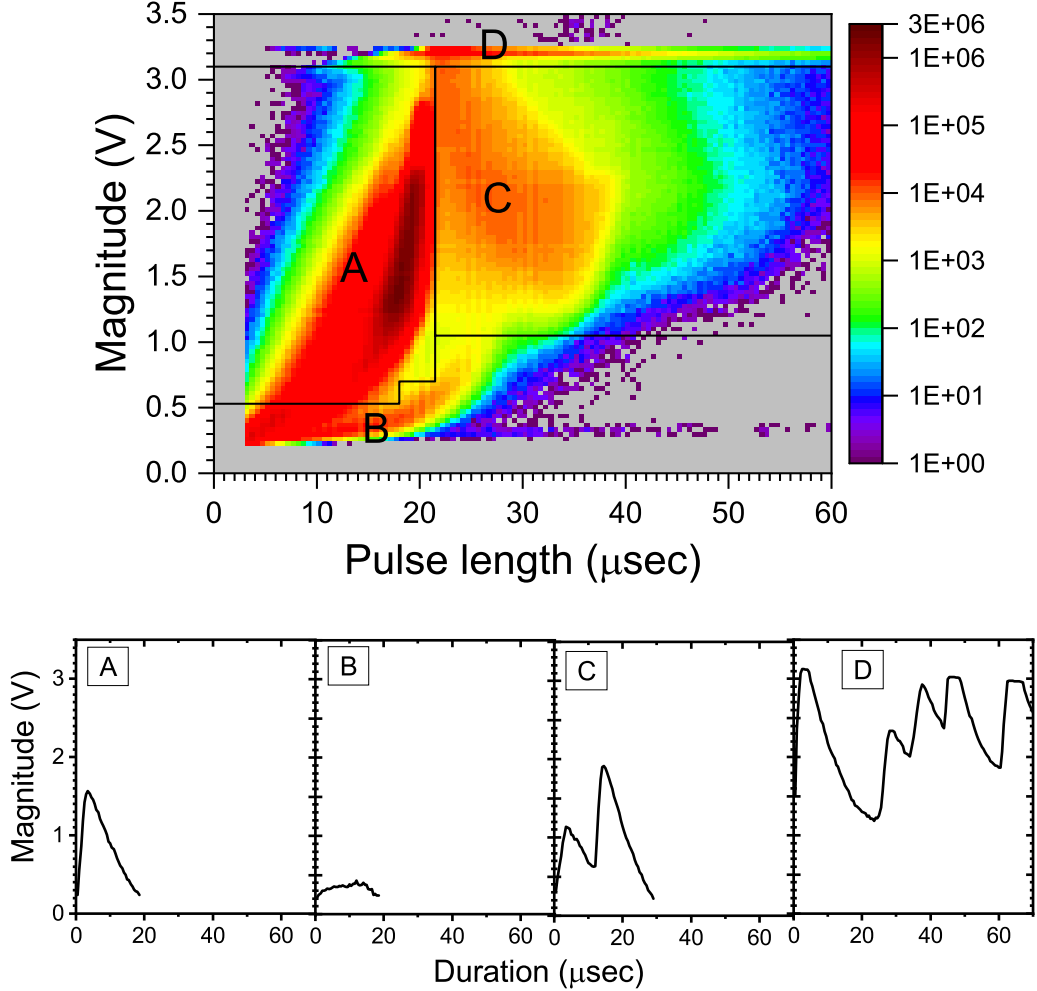}
\caption{
 Upper panel: 2D histogram of the {magnitude} vs. length of pulses recorded by DOMC NM (about $3\cdot$$10^8$ pulses)
 during the analyzed period (no additional threshold applied).
 Letters denote different branches as discussed in the text and summarized in Section~\ref{Sec:pulse}.
 Separation between the branches is shown by solid lines.
 Bottom panel: Typical profiles of pulses from different branches shown in the upper panel.
 {Panel D depicts an extremely long multiple pulse, while a typical pulse shape of
  branch D is shown in Figure~\ref{Fig:map_DOMB}D.}
  }
\label{Fig:map_DOMC}
\end{figure}

\section{DOMC/DOMB DAQ electronic system}

The new DAQ system is built with a single-board computer Raspberry Pi 3B and
 easily replaceable modules responsible for the operation of different subsystems
 (high voltage power supply, pre-amplifier and detector signal processor, temperature-pressure-humidity sensors
 and others) -- see full details in \citeA{strauss20}.

Pulse signals coming from the detector are amplified and digitized by a signal registration board.
It has a dedicated microcontroller PIC32 and built-in 10-bit analogue-to-digital converter (ADC)
 with the reference voltage set at 3.3 V.
Each pulse is sampled at the frequency of 2 MHz (viz. 0.5 $\mu$sec),
 and the information about its {magnitude} time profile is stored in a buffer of the board.
The central computer has software-defined discriminators in pulse's {magnitude} and length.
If a pulse in the buffer matches both criteria, it gets recorded in a data file by the central computer.
The files are compressed and sent to the data server of Oulu cosmic ray station
 for further import to databases \url{cosmicrays.oulu.fi} and \url{nmdb.eu}.

\begin{figure}[t]
\centering
\includegraphics[width=\columnwidth]{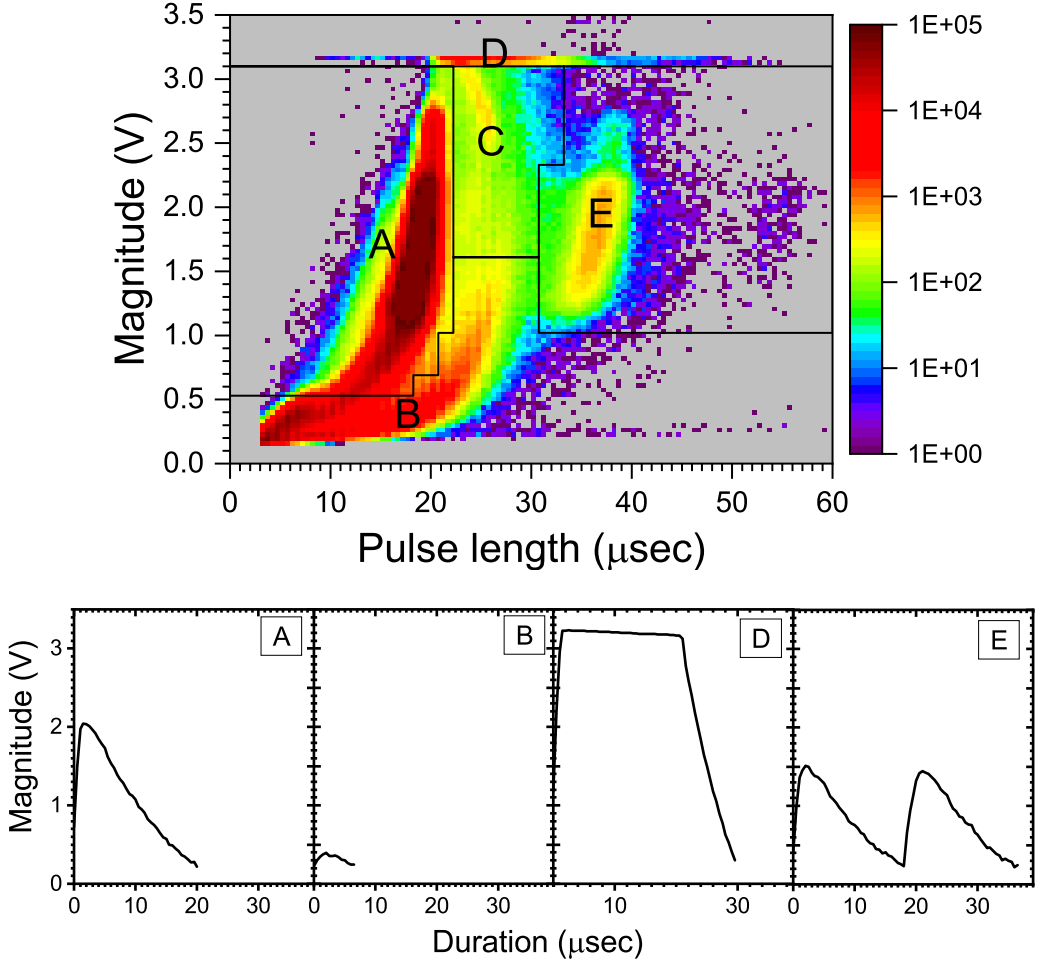}
\caption{
 The same as in Figure~\ref{Fig:map_DOMC} but for the DOMB NM (about $3\cdot$$10^8$ pulses).
  }
\label{Fig:map_DOMB}
\end{figure}
DOMC and DOMB have the following default thresholds for registered pulses:
 the {magnitude} discriminator at 0.2 V and the minimum pulse length of five sample points (2.5 $\mu$sec).
All pulses failing to meet these criteria are ignored by the DAQ system.
Since each pulse with the {magnitude} exceeding 0.2 V {and longer than 2 $\mu$sec}
 is digitized, a higher-value thresholds can be applied electronically in the off-line analysis.

Analyses of the pulse shapes and statistics are presented in the subsequent sections.

\section{Analysis of pulse shapes}
\label{Sec:pulse}

Here, we used data from DOMC for 01-Jan through 31-May-2020 (about $3\cdot10^8$ pulses recorded),
 and from DOMB for 13-Aug through 31-Oct-2019 ($4\cdot10^7$ individual pulses recorded).
These periods correspond to very quiet solar conditions (solar cycle minimum) with relatively low heliospheric modulation
 and high intensity of galactic cosmic rays (GCR).
There were no solar particle events or other notable transients during the studied period.
{The period of January through May 2020 corresponded to a transition from polar-day (around-a-clock
 insolation) to polar-night (no sunlight) conditions and formation of the polar vortex.
The period of August\,--\,October 2019 was characterized by very stable and cold weather with a stable
 polar vortex.
An example of temporal variability of the {pressure-corrected} count rates in different branches
 (see below) are shown in Figure~\ref{Fig:count_rates} for the first 42 days of each analyzed period.
{For correction, the following barometric coefficients were used: -0.769 \%/hPa for DOMC and -0.754 \%/hPa for DOMB
 (the reference pressure level 650 hPa) as defined during the operation of the detectors (see, e.g., metadata in http://cosmicrays.oulu.fi).}
One can see that the overall pressure-corrected count-rate vary within $\pm 1$\% only,
 but the variability in other branches is greater as described below.}
\begin{figure}[t]
\centering
\includegraphics[width=0.9\columnwidth]{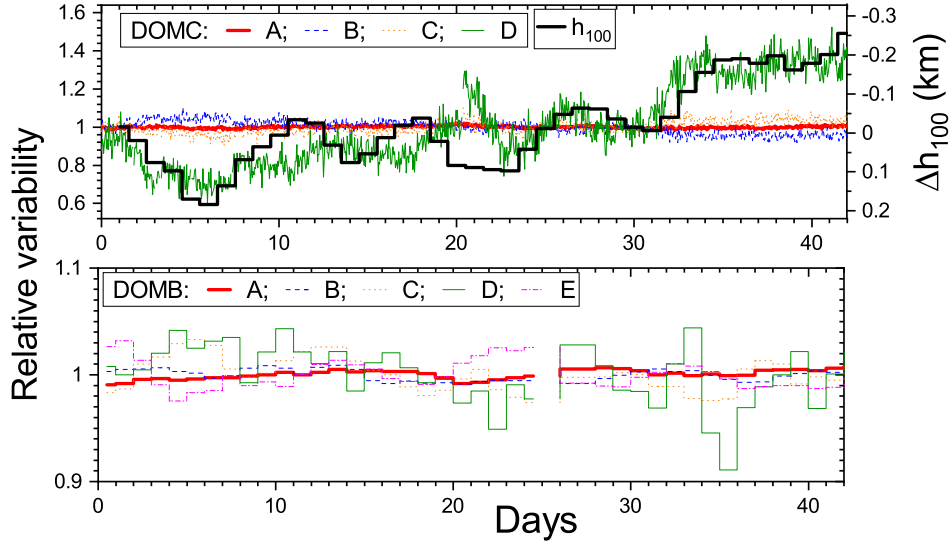}
\caption{
Pressure-corrected and normalized per unity count rates of NMs in different branches (see text) for the first 42 days of each NM dataset.
Upper panel: Hourly values for DOMC for the period 01-Jan through 11-Feb-2020-
{The thick black line depicts the anomaly, from the mean level 15.6 km, of the geopotential
 100 hPa height $h_{100}$ (right-hand Y-axis) over the Concordia station (ERA-5 data from ECMWF
 https://www.ecmwf.int/en/forecasts/datasets/reanalysis-datasets/era5).
Lower panel: Daily means for DOMB for the period 13-Aug through 25-Sep-2019.
}  }
\label{Fig:count_rates}
\end{figure}

For the analysis, we used the following pulse parameters: {magnitude} $A$ (the maximum voltage); length $t$ (the duration, in $\mu$sec
 of the signal being above the selected threshold); the {$e$-folding} decay time $\tau$ of the exponential decline of the pulse voltage
 after the maximum, and the waiting time $\Delta T$ between onsets of subsequent pulses (in $\mu$sec).

First, we analyzed the relation between the {magnitude} $A$ and the length $t$ of all individual pulses.
These distributions are presented in Figures~\ref{Fig:map_DOMC} and \ref{Fig:map_DOMB}, respectively for the
 DOMC and DOMB NMs \cite<cf. Figure 4 in>[]{strauss20}.
Several clearly distinguished branches can be identified (see statistic in Table~\ref{Tab:branch_b}):

\subsection{Branch A: Normal pulses}
The main branch contains the majority ($>$90 \%) of the pulses.
It consists of single, well-defined pulses with a fast rise (a few $\mu$sec) followed by an exponential
 decline with the {$e$-folding decay time $\tau\approx 8$ $\mu$sec, defined by the pre-amplifier's circuit
  (relaxation of a capacitor).}
Samples of typical pulses in this branch can be observed in Figures~\ref{Fig:map_DOMC}A and \ref{Fig:map_DOMB}A
 for DOMC and DOMB NMs, respectively.
The {magnitude} takes {the entire range up to 3 V}, and duration 5\,--\,20 $\mu$sec, with a tendency
 that higher pulses are slightly longer, as they decay to the detection threshold level longer.
{However, since low pulses ($A$$<$0.5 V) are mixed with the branch B (noise), we consider normal pulses
 as those with $A$$>$0.5 V.
The time variability of the count rate in this branch is perfectly corrected
 for pressure {(the formal Pearson correlation between the daily count rate and the pressure is $r=$-0.02)}
  and remains stable within $\pm 1$\% in both DOMC and DOMB.}
This is the clear signal part which forms the main fraction ($\approx 91$\%) of the count rate, while the remaining
 8\,--\,9 \% of pulses need more discussion.

\subsection{Branch B: Noise}
This branch contains very low ($A<0.6$ V) pulses without any clear shape (see Figures~\ref{Fig:map_DOMC}B and
 \ref{Fig:map_DOMB}B).
The positive correlation between $A$ and $t$ is expected for the same reason as for branch A.
These pulses are likely related to electronic noise.
This component comprises 5\,--\,7 \% of the total number of pulses but can be reduced to $<$1\%,
 {thus improving the signal-to-noise ratio from $\approx$15 to $>$90}
 by increasing the {magnitude} threshold level up to 0.5\,--\,0.6 V.
{Since the short-length ($<$12 $\mu$sec) pulses are contributed also from branch A (normal pulses),
 the percentage above is a conservative upper limit.
This branch depicts hardly any time variability as expected for the noise, but after nominal pressure correction
 (Figure~\ref{Fig:count_rates}) it appears 'over-corrected' and depicts slight variability in phase with the pressure
 {(r=0.29$\pm$0.15, $p-$value =$0.03$).}

\subsection{Branch C: Contribution of atmospheric cascade}
This branch consists of moderately high ($A>1.5$ V) and longer (20\,--\,35 $\mu$sec) pulses.
They are typically double peaks (Figure~\ref{Fig:map_DOMC}C) where the second peak starts before the first one
 drops below the detection level.
The negative correlation between the {magnitude} and the length is understandable, as the second pulse starts over non-zero background.
The separation between the pulses is from 5\,--\,20 $\mu$sec.
The sub-pulses are totally consistent with the pulses in branch A in duration and decay time ($\tau\approx 8\, \mu$sec).
This branch is clear in DOMC data, comprising about 3\% of all pulses, but is hardly visible (only 0.24 \%)
 in the DOMB dataset, {implying that is caused by a nucleonic component.}
The time separation between the subpulses ($<20\, \mu$sec) is much longer than the characteristic time (expected to be of the order
 of several nanoseconds) of a cascade within the detector itself, including lead producer, but shorter than the full development
 of the atmospheric cascades (see Section~\ref{sec:WTD}).
It is likely related to a {tail of the WTD for the atmospheric cascade development (Section~\ref{sec:WTD}), when the time separation
 between secondary nucleons of the same cascade appear shorter than the single pulse length of $\approx$ 20 $\mu$sec.
Such pulses are registered as long ones composed of two partly overlapping pulses and form branch C.
Contributions of nucleons with longer time separation make single pulses associated to branch A.}
The count-rate variability in this branch (Figure~\ref{Fig:count_rates}) still depicts dependence {(under-correction)}
 on pressure after the nominal pressure correction  {(r=-0.26$\pm$0.16, $p$=$0.05$}), implying a stronger dependence
 on pressure than that for branch A.
We note that the standard NM64 electronic setup with the dead-time (20 $\mu$sec) {makes this branch indistinguishable from branch A}.

\subsection{Branch D: Possible contribution of muons}
This branch is characterized by very high {magnitude} ($A>3$ V, viz. near the upper bound of ADC)
 and long duration of pulses.
Most numerous here are saturated pulses (Figure~\ref{Fig:map_DOMB}D) but there are also very long pulses,
 up to 110 $\mu$sec in length, that include a sequence of short but very high sub-pulses following each other
 by several $\mu$sec (Figure~\ref{Fig:map_DOMC}D).
The saturated pulses would require, if fitted with a `standard' pulse shape (Fig.~\ref{Fig:map_DOMC}A), a very high magnitude of up to 30 V,
 viz. a factor of 10 greater than normal pulses.
Multiple pulses may contain up to 10 short sub-pulses, also implying an enhanced yield.
This points to a different type of process producing pulses in branch D.
The most plausible candidate is the process of direct multiple-ion production by a cascade muon in BF$_3$ gas inside
 the NM proportional counter \cite{knoll10,siciliano12}.
This can ignite multiple, nearly simultaneous electromagnetic avalanches inside the counter, leading to more `energetic'
 recorded pulses (voltage exceeding the upper limit and/or multiple overlapping pusles -- see Figures~\ref{Fig:map_DOMB}D
 and \ref{Fig:map_DOMC}D, respectively).
{This process has not been considered nor properly modelled for NMs, where its contribution is small, in contrast
 to the usually considered muon contribution to NM counts, viz. muon-induced production of neutrons
 in the lead producer \cite{clem00} with subsequent detection in the counter \cite{maurin15,mangeard2}}.
Such muon-induced neutrons are detected in a usual way and cannot be distinguished from the signal of the hadronic component.
Accordingly, they appear in branch A and cannot contribute to branch D.
Therefore, we can speculate that the branch D is likely related to non-hadronic particles producing abnormally energetic pulses
 via direct ionization of the filling gas.
{A more detailed study of this process is planned for the future.}

{
Count-rate in this branch (Figure~\ref{Fig:count_rates}) depicts strong variability: $\pm$25\% for DOMC
 and $\pm$5\% for DOMB, which can be explained by a strong change in the atmospheric density profile and by a stable
 vortex conditions, during the two analyzed periods, respectively.
For comparison, we shown in Figure~\ref{Fig:count_rates}A the temporal variability, for the same period,
 of the anomaly of the geopotential height of the $\Delta h_{100}$ (100 hPa) atmospheric level, which roughly corresponds to the
 mean height of the muon production and affects the muon flux near ground.
The branch-D count rate co-varies in sync with $\Delta h_{100}$ ($r$=-0.84, $p$-value $<10^{-6}$), and  the
 magnitude of the muon-flux variability is consistent with the $\Delta h_{100}=0.5$ km using the measured spectrum of muons
 \cite{boezio_mu03} at a 100-hPa level and relativistic time dilation.
This confirms the muon origin of this branch since no other source can reliably explain it.
We emphasize that this effect is strong for the high altitude of DOMC location but fades towards lower heights (because
 of the higher energy of muons that can reach lower altitudes) and accounts for only 5\% at sea level of 1013 hPa.
Thus, the NM with a high {magnitude} threshold ($V_0>3.1$ V) can operate as a low-efficiency muon detector.
}

\subsection{Branch E: Possible contamination from DOMC}
There is a branch with pulses of a normal {magnitude} ($A$ between 1\,--\,2.5 V) and double length
 (35\,--\,40 $\mu$sec) clearly visible for DOMB NM (Figure~\ref{Fig:map_DOMB}).
This branch contains 0.28~\% of all pulses and is composed of double pulses (Figure~\ref{Fig:map_DOMB}E)
 separated by 10\,--\,20 $\mu$sec.
Interestingly, this branch is not distinguishable in DOMC data (Figure~\ref{Fig:map_DOMC}).
We do not have a clear understanding of the origin of this branch but may speculate
 that it is possible contamination from {multiple neutrons produced by the lead producer
 of the neighbouring DOMC detector.
Because of the diffusive propagation, the neutrons may arrive to DOMB at slightly different times.
The detectors are located about 1 meter apart of each other, leading to a 20-$\mu$sec traversing time
 for one meter of air and several g/cm$^2$ of the moderator/reflector layers.
We note that if the two pulses are separated by more than 25 $\mu$sec so that the voltage drops below
 the discriminator's level, the pulses are counted as two separate ones.
An insignificant hint on triple pulses can be observed in Figure~\ref{Fig:map_DOMB} at 50 $\mu$sec.
We note that 20 $\mu$sec is known as electronic `dead-time' of a standard NM \cite{hatton64}.
Nothing conclusive can be seen in the time variability of this branch, because
 of the low statistics (about 1 count per minute).
}

{This branch could be potentially caused also by noise in the pre-amplifier's electric circuit,
 by producing an `echo' of the signal with a delay time of $\approx$20 $\mu$sec, viz. double\,--\,triple
 characteristic time of the circuit (8 $\mu$sec).
However, this is unlikely since such echoing was not observed during electronic tests of the DAQ board.}

\begin{table}
\caption{Different branches of pulses ({magnitude}-vs-length) of the DOMC and DOMB detector
 (see Figures~\ref{Fig:map_DOMC} and ~\ref{Fig:map_DOMB}, respectively).
 Percentages for the branches were computed using the boundaries indicated in Figures~\ref{Fig:map_DOMC} and \ref{Fig:map_DOMB}. }
\label{Tab:branch_b}
\begin{tabular}{c|cc|p{9cm}}
\hline
Branch & \multicolumn{2}{c|}{percentage} & Possible origin \\
 & DOMC & DOMB & \\
\hline
A & 90.98 & 91.49 & Main branch: single pulses \\
B & 5.20$^\dagger$ & 7.84$^\dagger$  & Noise\\
C & 3.15 & 0.23 & Multiple pulses from the same atmospheric cascade\\
D & 0.68 & 0.16 & {Possible muon contribution} \\
E & -- & 0.28 & Possible contamination from the neighbouring detector\\
\hline
\end{tabular}
\\ {$^\dagger$ the percentage is a conservative upper bound as this branch includes also pulses from branch A.}
\end{table}

\section{Waiting time distribution}
\label{sec:WTD}
\begin{figure}[t!]
\centering
\includegraphics[width=0.8\columnwidth]{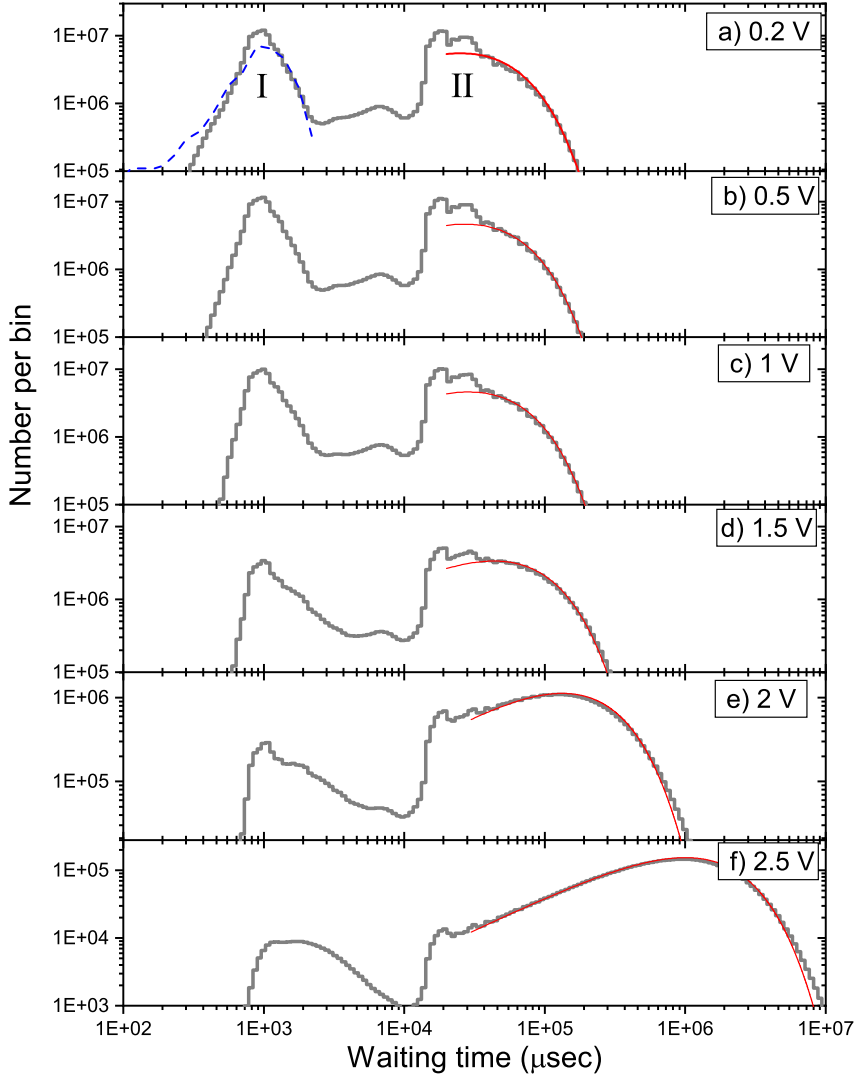}
\caption{
 Logarithmically binned histograms of the waiting-time ($\Delta T$) distribution of pulses recorded by DOMC NM for different
  discriminator's {magnitude} threshold levels $V_0$, as indicated (in volts) in the legends.
  Red curves represent the theoretically expected distribution for randomly occurring independent events, corresponding to the count rate.
  The blue dashed curve in panel a) represents a simulated delay of the arrival time of $\leq$10 keV neutrons with respect to the
   cascade-front arrival for a 10 GeV cosmic-ray proton.
  }
\label{Fig:WT_DOMC}
\end{figure}
Next, we have analyzed the waiting-time distribution (WTD) of recorded pulses.
We define the waiting time $\Delta T$ as the time interval between the onsets of consecutive pulses recorded
 by the DAQ system of a NM.
Since the length of individual pulses can reach 100 $\mu$sec (Figures~\ref{Fig:map_DOMC} and \ref{Fig:map_DOMB}),
 we analyse the WTD for $\Delta T>100\, \mu$sec.
The observed WTDs are plotted, as logarithmically-binned histograms in Figures~\ref{Fig:WT_DOMC} and \ref{Fig:WT_DOMB},
 respectively for DOMC and DOMB NMs, for different values of the {magnitude} threshold $V_0$ so that $A\geq V_0$.
The distributions for different values of $V_0$ have a similar shape with two clearly separated peaks:
 one (called peak $I$) at $\Delta T \approx 1$ msec, the other (peak $II$) located between 30 msec and 10 sec
 depending on the value of $V_0$ and the detector type.

\subsection{Peak I: In the atmospheric cascade}

The first WTD peak (peak $I$) is located at $\Delta T\approx 1$ msec.
Its location is very stable and does not depend on the $V_0$ values, nor on the detector type.
On the other hand, the peak broadens for higher threshold values extending its tail to longer waiting times,
 up to several milliseconds.
The contribution of this peak to the total count rate varies from about 35\% (no additional threshold) down to 2.5\%
 ($V_0=2.5$ V) for DOMC NM, and from 7\% down to 0.1\% for the DOMB NM, respectively.
We note that, while the bulk of secondary protons, muons, and electromagnetic components of the cascade
 arrive to a detector as a relatively thin front, secondary neutrons diffuse in the atmosphere, leading to
 a wide spread in time and lateral distribution compared to other secondaries \cite{grieder11}.
\begin{figure}[t!]
\centering
\includegraphics[width=0.8\columnwidth]{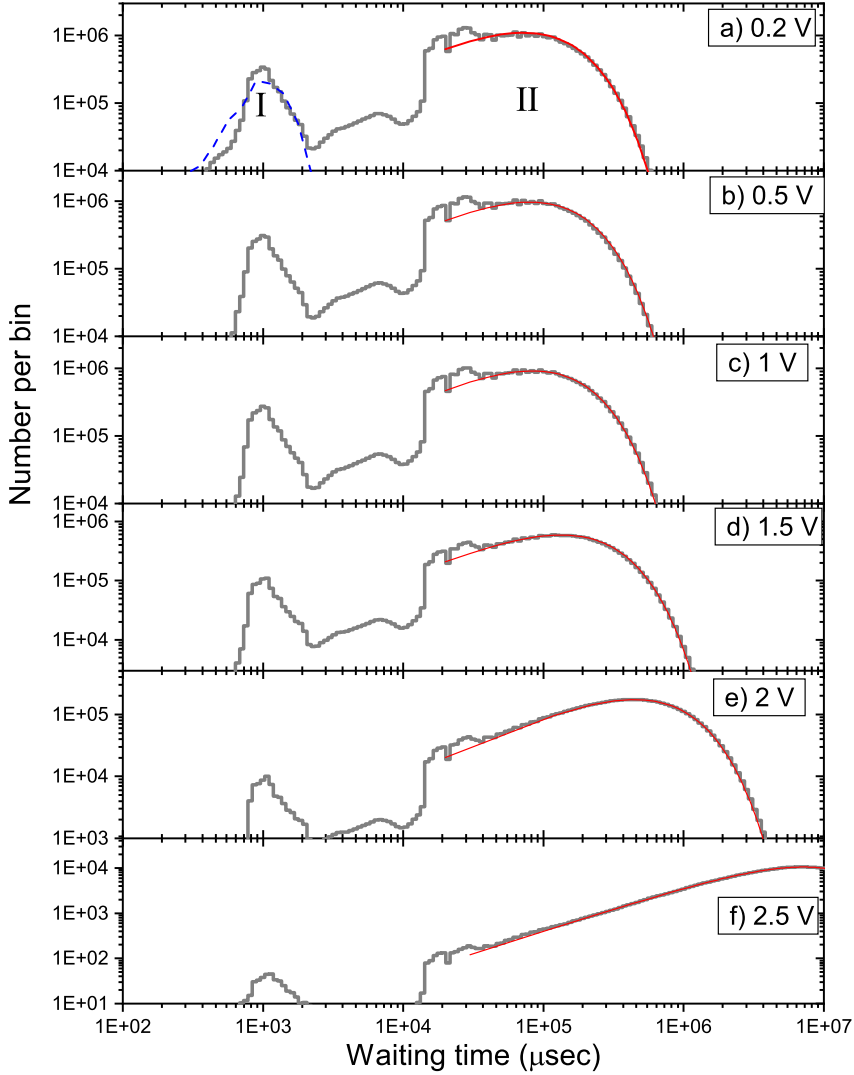}
\caption{
 The same as Figure~\ref{Fig:WT_DOMC} but for DOMB NM.
  }
\label{Fig:WT_DOMB}
\end{figure}

In order to check that, we performed a numerical simulation of the development of the atmospheric cascade
 using the Monte-Carlo simulation toolbox Geant4 v.10.6.0 \cite{agostinelli03, allison06}
 with the physics list QGSP\_BIC\_HP (Quark-Gluon String model,
 Geant4 Binary Cascade model, High-Precision neutron package) \cite{Geant4-Phys2020}.
We simulated $10^6$ atmospheric cascades initiated
 by incident protons with the kinetic energy of 10 GeV impinging vertically on the top of the atmosphere.
We note that this energy corresponds to the effective energy of a polar NM to GCR \cite{kudela00,asvestari_JGR_17}
 and thus roughly represents the relation between the NM count rate and CR variability.
During the simulations, we traced secondary neutrons and recorded their crossing of the reference
 level of 650 g/cm$^2$ where the DOMC/DOMB NMs are located.
For each neutron crossing, the neutron's energy, location with respect to the cascade axis and the time since
 the first interaction were recorded for further analysis.
Neutrons were found to spread as far as 6 km from the cascade axis \cite<see also>[]{paschalis11} with the delay of up to 80 milliseconds.
{Next, we built a logarithmically-binned histogram of the WTD between arrivals of epithermal neutrons ($\leq 10$ keV) with
 spatial separation less than 1 meter, from the same atmospheric cascade, as shown by the
 blue dashed curve in Figures~\ref{Fig:WT_DOMC}a and \ref{Fig:WT_DOMB}a}.
The distribution reasonably well matches both the location and width of peak I.
The height of the distribution was scaled up to match the observed WTD, while keeping the shape and location
 of the peak.
Similar results can be obtained for other energies of the primary particle since the 1-msec time is caused
 by diffusion and thermalization of secondary neutrons from 1 MeV (evaporation peak) to 10 keV energy,
 and not by the development of a cascade per se.

Accordingly, WTD peak I can be reliably associated with the neutron diffusion and thermalization
 within atmospheric cascades, and the time of about one millisecond is a typical time-scale for such a process.
These pulses generally contribute to the well-known multiplicity of NM counts \cite{debrunner68a,balabin11,mangeard2,ruffolo16},
 viz. multiple correlated pulses within the NM count rate.
The mean multiplicity of the NM count rate for different threshold $V_0$ values is shown in Figure~\ref{Fig:mult}.
It is calculated as the ratio of the total number of counts for each NM to that in peak II, viz. the number of
 individual cascades.
The obtained values for the multiplicity (1.5 and 1.06 for $V_0=0$ for DOMC and DOMB, respectively) are
 slightly higher than those measured at sea level \cite{hatton64} but lower than that for an air-borne NM
 \cite{kent68}.
\begin{figure}[t!]
\centering
\includegraphics[width=0.7\columnwidth]{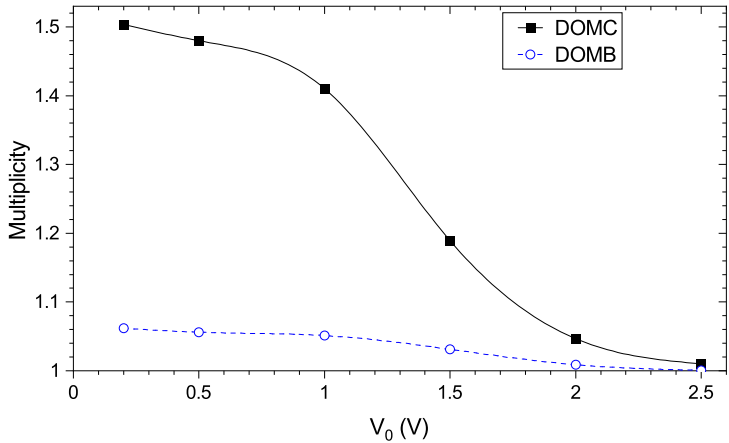}
\caption{
 Mean multiplicity of the DOMC {(black squares)} and DOMB {(blue open circles)} NMs as a function of the threshold $V_0$.
  }
\label{Fig:mult}
\end{figure}

\subsection{Peak II: Individual atmospheric cascades}

The other peak (called peak II) is quite broad and corresponds to the waiting times from about 20 msec to several seconds.
{The distribution is slightly distorted between 10 and 40 msec, probably due to interference with the
 power-line frequency (50 Hz).}
The peak has a well-defined smooth shape, and its location depends on the detector (DOMB and DOMC) and the
 value of the threshold $V_0$ (Figures~\ref{Fig:WT_DOMC} and \ref{Fig:WT_DOMB}).
This peak in WTD corresponds to individual atmospheric cascades caused by the primary cosmic-ray particles.
In order to illustrate this, we have also plotted (as red curves) the analytically expected WTD for
 randomly occurring independent pulses with the occurrence probability defined by the observed count rate
 $\nu$ of a given NM (DOMC or DOMB) for a given value $V_0$.
WTD of independently occurring events with the occurrence probability $\nu$ is expected to be exponential with the characteristic
 decay time $1/\nu$, which takes for a logarithmically binned histogram the shape shown by the red curves.
One can see that the analytical WTD perfectly describes the observed peak II for different values of $V_0$ and
 for both DOMC and DOMB, confirming its relation to the individual atmospheric cascades.

The observed WTD is consistent with the recently introduced delay time--geometry correction factor for the NM
 yield function, details are given elsewhere \cite{mishev13, mishev20}.

\section{Discussion and conclusions}

The new DAQ system of neutron monitors makes it possible to study different processes induced by
 cosmic rays in the atmosphere and the detector itself.
In Sections~\ref{Sec:pulse} and \ref{sec:WTD} we have analyzed different pulse parameters and
 waiting times.
Somewhat similar analyses were made also earlier \cite{hatton68} using oscilloscopes and were
 based on small statistic.
The new DAQ system allows continuous analysis of the pulses, e.g., the statistic shown here
 includes $3\cdot 10^8$ and $4\cdot 10^7$ individual pulses for the DOMC and DOMB, respectively,
 for illustration, but it can be much greater.
Thanks to the fully digitized pulses, the same dataset recorded by the new DAQ system allows
studying different processes separately.

Individual atmospheric cascades can be studied as pulses separated by more than two milliseconds, which is
 close to the standard NM detection mode with the dead-time of 1.2 msec \cite{hatton64}.
It is important that, in contrast to the standard NM64 DAQ system with the fixed ``dead-time'',
 waiting time between pulses can be now performed by software data-processing.

Development of the atmospheric cascade can be studied using pulses with waiting times between 0.3 and 3 msec.
This is not usually done in the NM data analysis directly but rather via the multiplicity (or `leader-fraction') analyses.
With the new DAQ system, one can study details of the cascade development for each cascade individually,
  including also the multiplicity.

Development of the {atmospheric cascade}
 in the vicinity of the detector can be studied using the pulse-shape analysis, in
 particular, in branches C and D in Figures~\ref{Fig:map_DOMC} and \ref{Fig:map_DOMB}.
In particular, with the high {magnitude} threshold, the NM can operate as a low-efficiency muon detector.
We are not aware of any similar analysis done previously.
Moreover, a detailed analysis of branches C and D allows one to discriminate hadrons from muons, and hence to use an
 NM in a regime of muon detector, using the same dataset.

The noise is shown to form a well-separated branch, being characterized by the low pulse {magnitude}.
Accordingly, by selecting the discriminator threshold to the value of $V_0 =$ 0.5\,--\,0.6 V (the default is 0.2 V)
 one can cut off the noise effectively without a significant reduction of regular pulses.
Placing the threshold to 0.5 V eliminates about 90\% of the noise and only a few percent of normal pulses,
 thus increasing the signal-to-noise ratio by an order of magnitude.

The standard NM is an energy-integrating detector and cannot measure the energy spectrum of cosmic rays.
Combining NM with different geomagnetic cutoff rigidities to the worldwide network makes them a very rough spectrometer \cite{moraal00}.
It works reasonably well for GCR \cite{caballero12} but cannot be directly applied to a study of
 solar energetic particles because of the
 possible anisotropy which can be large during impulsive (phase of) events \cite<e.g.,>[]{mishev14}.
Data from different NMs cannot be compared directly without a complicated analysis of magnetospheric
 transport of charged particles \cite<e.g.,>[]{smart00}.
The use of a pair of standard and bare NMs in the same location (e.g. at the South Pole, SANAE or Dome C stations)
 provides a rough measure of the hardness of the cosmic-ray spectrum \cite{caballero16,nuntiyakul18}, based on the ratio of their count rates.
Here we propose that the use of different $A$-values in the new DAQ-system dataset may
 provide an estimate of the cosmic-ray spectrum.
Specifically, a muon-related branch of the pulses can be separated, providing muon counts in the same detector.
Fine analysis of relatively weak ground-level solar particle events is often very limited due to the shortage of
 information about the energy spectrum since such events are seen only by a few high-altitude polar NMs with no significant response
 from sea-level and non-polar instruments \cite{mishev17}.
In such a case, any spectral information is very crucial and any new addition significantly increases the quality of the analysis.
Varying the value of $A$, one can obtain several spectral points.
In particular, the measured waiting time and {magnitude} distributions of pulses could serve for that purpose.
In an ideal case, a strong SEP event, such as the GLE\#69 of 20-Jan-2005, could be used to `calibrate' the detector,
 but such events occur very seldom.
Alternatively, this possibility can be explored quantitatively with a full simulation of
 the detector-and-atmosphere response to the primary particles.
This is doable with the modern Monte-Carlo simulation techniques, but is beyond the framework of this paper and left for
 forthcoming work.

In summary:
\begin{enumerate}
\item A new DAQ system has been installed on DOMC and DOMB NMs that has gathered a large amount of fully digitized, at a sub-$\mu$sec
 sampling rate, individual pulses.
\item An analysis of the data has demonstrated clustering of pulses to several branches: (A) the main branch representing secondary neutrons;
  (B) detector's electronic noise; (C) double pulses caused by {shortly separated pulses of the same} atmospheric cascade;
  (D) multiple pulses likely related to atmospheric muons;
  (E) possible contamination of DOMB detector by neutrons scattered from the neighbouring DOMC detector.
\item An analysis of the waiting time distributions has revealed two clearly distinguishable peaks: peaks I at about 1 msec related to
 the intra-cascade diffusion and thermalisation of secondary atmospheric neutrons; and peak II (30\,--\,1000 msec) corresponding to individual atmospheric cascades.
\item
It is shown that a NM with the new DAQ system can provide also data on muon flux, using the same dataset.
\end{enumerate}
This opens a new possibility to study spectra of cosmic-ray particles in a single location and details of cosmic-ray induced
 atmospheric cascades.

\acknowledgments
We are grateful to the personnel of Concordia station hosting the DOMC/DOMB instrumentation.
Operation of DOMC/DOMB NMs is possible thanks to the hospitality of the Italian polar programme PNRA (via the
 LTCPAA PNRA 2015/AC3 and the BSRN PNRA OSS-06 projects) and the French Polar Institute IPEV.
It is supported by the Academy of Finland (projects CRIPA-X No. 304435, ESPERA No. 321882, QUASARE 330063, HEAIM-2 330427),
 and Finnish Antarctic Research Program (FINNARP).
DOMC/DOMB NM data can be obtained from http://cosmicrays.oulu.fi, courtesy of the Sodankylä Geophysical Observatory.


%
%

\bibliography{usoskin_all}

%
%
%
%
%

\end{document}